\documentclass[%
preprint,
 amsmath,amssymb,
 aps,
]{revtex4-1}

\usepackage{graphicx}
\usepackage{dcolumn}
\usepackage{bm}

\newcommand{\rbar}[1]{\left.#1\right|}
\def\e{\mathrm{e}}
\def\cc{\mathrm{c.c.}}
\def\Re{\mathrm{Re \,}}
\def\Im{\mathrm{Im \,}}
\def\ii{\mathrm{i}}

\begin{document}

\title{Boundary streaming with Navier boundary condition}

\author{Jin-Han Xie}
 \email{J.H.Xie@ed.ac.uk}
\author{Jacques Vanneste}%
\affiliation{%
School of Mathematics and Maxwell Institute for Mathematical Sciences, University of Edinburgh, Edinburgh EH9 3JZ, UK
}%

%
%

\date{\today}

\begin{abstract}
In microfluidic applications involving high-frequency acoustic waves over a solid boundary, the Stokes boundary-layer thickness $\delta$ is so small that some non-negligible slip may occur at the fluid-solid interface. 
This paper assesses the impact of this slip by revisiting the classical problem of steady acoustic streaming over a flat boundary, replacing the no-slip boundary condition with the Navier  condition $u|_{y=0} = L_\mathrm{s} \partial_y u|_{y=0}$, where $u$ is the velocity tangent to the boundary $y=0$, and the parameter $L_\mathrm{s}$ is the slip length.
A general expression is obtained for the streaming velocity across the boundary layer as a function of the dimensionless parameter $L_\mathrm{s}/\delta$. 
The limit outside the boundary layer provides an effective slip velocity satisfied by the interior mean flow. Particularising to travelling and standing waves shows that the boundary slip respectively increases and decreases the streaming velocity.
\end{abstract}

\maketitle

\section{Introduction}

Among the many techniques devised to manipulate fluids at microscales \citep[e.g.][]{Squi2005, Tabe2005}, the use of high-frequency acoustic waves appears particularly promising. As a result, the field of what \citet{Frie2011} term acoustic microfluidics is rapidly expanding; see Ref.\ \citenum{Frie2011,Yeo2013} for reviews of the experimental and theoretical state of the art in this field. 

One of the main ingredients in the techniques developed is streaming---the generation of mean flow by dissipating acoustic waves. Two forms of streaming can be distinguished  \citep{Ligh1978,Rile2001}: (i) interior streaming, induced by wave attenuation in the fluid interior \citep{Ecka1948,West1953,Nybo1953,Nybo1965}; and (ii) boundary streaming \citep{Rayl1896} which is confined near solid boundaries but influences the interior mean flow by modifying its effective boundary condition \citep[see also][]{Long1953,Brad1996}. Both types of streaming share the remarkable property of non-vanishing mean motion in the limit of vanishing viscosity \citep{Nybo1965,Ligh1978}; both contribute to the interior mean flow, although the boundary contribution is small when the acoustic wavelengths are small compared to the flow scales \citep{Vann2011}.

A feature of many experiments in acoustic microfluidics \citep[e.g.][]{From2008,Roge2010,Algh2011,Mano2012,Dent2014} is the high frequencies employed. A consequence is that the Stokes boundary-layer thickness  is very small. This thickness estimates the size of the near-boundary region where viscous effects dominate and is given by $\delta=\sqrt{2\nu/\omega}$, where $\nu$ is the fluid's kinematic shear viscosity and $\omega$ is the wave's angular frequency. In water, and for typical frequencies in the range 1 MHz to 1 GHz, $\delta$ is in the range 500 nm to 10 nm. This implies large stresses at the fluid-solid interface and, as a result, suggests that the no-slip boundary condition that is traditionally used for the study of boundary streaming may not be appropriate \cite{Laug2007}. 

Motivated by this observation, we assess the effect that the possible slip of the fluid along the boundary has on boundary streaming. We do so by revisiting the classical model of boundary streaming over a flat plate, replacing the no-slip boundary condition by the more accurate Navier boundary condition \citep{Navi1823}
\begin{equation} \label{navier}
u|_{y=0} = L_\mathrm{s} \rbar{\partial_y u}_{y=0},
\end{equation}
where $y=0$ defines the boundary, $u$ is the velocity tangent to the boundary, and $L_\mathrm{s}$ is the so-called slip length, a property of the fluid-solid interactions \citep[e.g.][]{Laug2007,Tabe2005}.
The key dimensionless parameter in the problem is the ratio
\begin{equation}
\beta = L_\mathrm{s} / \delta
\end{equation}
of the slip length to the Stokes boundary-layer thickness. With typical values for $L_\mathrm{s}$ of 10 to 100 nm (see e.g.\ Ref.\ \citenum{Laug2007}), this parameter can take a broad range of values.

We examine the streaming induced on a motionless flat boundary by a plane acoustic wave in the far field.
This is a simple problem, which we solve explicitly using a matched asymptotics technique relying on the small parameter $\delta k$, where $k$ denotes the acoustic wavenumber. The solution is instructive, however, since the effect of slip,
$\beta \not=0$,
on the streaming velocity is not obvious a priori: on the one hand, the slip reduces 
the shear and hence the Reynolds stress associated with the wave field; on the other hand, by weakening the constraint at the wall, it can increase the mean flow response to a given wave forcing. The non-trivial impact of the slip is illustrated by the fact that travelling and standing waves---two particular cases of our more general set-up---have different responses, respectively an increase and a decrease of the streaming velocity outside the boundary layer as $\beta$ increases from zero.

\section{Wave field}

We consider a plane acoustic wave with velocity
\begin{equation} \label{U1}
\mathbf{U}_1 = \Re \left(U(x) \e^{- \ii\omega t} \mathbf{e}_x \right),
\end{equation}
propagating over a horizontal plate located at $y=0$. Here $U(x)$ is an arbitrary complex function, $\omega$ is the (angular) frequency and $\mathbf{e}_x$ the unit vector in the $x$ direction.  
Note that the form (\ref{U1}) includes both travelling waves (for which $U(x) \propto \e^{ \ii k x}$) and standing waves (for which $U(x)$ is real).

The dynamics is governed by the compressible Navier--Stokes equations
\begin{equation}
\begin{aligned}
{\partial_t \rho} + \nabla \cdot \left( \rho \mathbf{u} \right) & = 0,\\
\rho {\partial_t \mathbf{u}}  + \rho \mathbf{u} \cdot \nabla \mathbf{u} &= -\nabla p + \mu \nabla^2 \mathbf{u} + (\mu^\mathrm{b} + \mu/3) \nabla \nabla \cdot \mathbf{u}, \label{NSeq}
\end{aligned}
\end{equation}
where $\mu$ and $\mu^\mathrm{b}$ are the shear and bulk viscosities, supplemented by an equation of state $p=p(\rho)$. Assuming that $U(x)$ is small compared with the sound speed $c_0$, we introduce the expansions
\begin{align}
\ \mathbf{u} &= \mathbf{u}_1 + \mathbf{u}_2 + \ldots, \\
\ p - p_0 &=  p_1 +  p_2 + \ldots, \\
\ \rho - \rho_0 &= \rho_1 +  \rho_2 + \ldots,
\end{align}
where the subscripts indicate the order in $U/c_0$. We are seeking a perturbative solution of (\ref{NSeq}) with $\mathbf{u}_1$ matching the far-field form (\ref{U1}) away from the boundary and satisfying the Navier boundary condition (\ref{navier}) at $y=0$. We consider the case of a small viscosity, characterised by $k \delta \ll 1$, with $k=\omega/c_0$ the wavenumber; in this case, the effect of viscosity is confined to a layer of thickness $\delta$ above the boundary. The solution in this boundary layer is best written in terms of the rescaled coordinate $Y=y/\delta$. This yields the order-one equations in the boundary layer,
\begin{equation}
\partial_t \rho_1+\rho_0 \left(\partial_x u_1+\delta^{-1}\partial_Y v_1\right) = 0,\label{021}
\end{equation}
which indicates that $ v_1/u_1 = O(k\delta) $, 
\begin{equation}
\rho_0 \partial_t {u_1} = - \partial_x p_1 +\mu \delta^{-2} \partial^2_{YY} u_1 \quad \textrm{and} \quad 
\partial_Y p_1 = 0, \label{014}
\end{equation}
where we have neglected terms of relative size $O(k \delta)$. Away from the boundary layer, in the outer region, the flow is irrotational and viscous terms are negligible, so $R_1 = \lim_{Y \to \infty} \rho_1$, $\mathbf{U}_1=\lim_{Y\to\infty} \mathbf{u}_1$ and $P_1 = \lim_{Y\to \infty} p_1$ satisfy
\begin{align}
\partial_t R_{1} + \rho_0 \partial_x U_1 &= 0, \label{005} \\
\rho_0 \partial_t \mathbf{U}_1 &= -\nabla P_1.
\end{align}
For consistency with (\ref{U1}), $V_1 = \lim_{Y \to \infty} v_1=0$.

\begin{figure*}
\centering
\includegraphics[width=14cm]{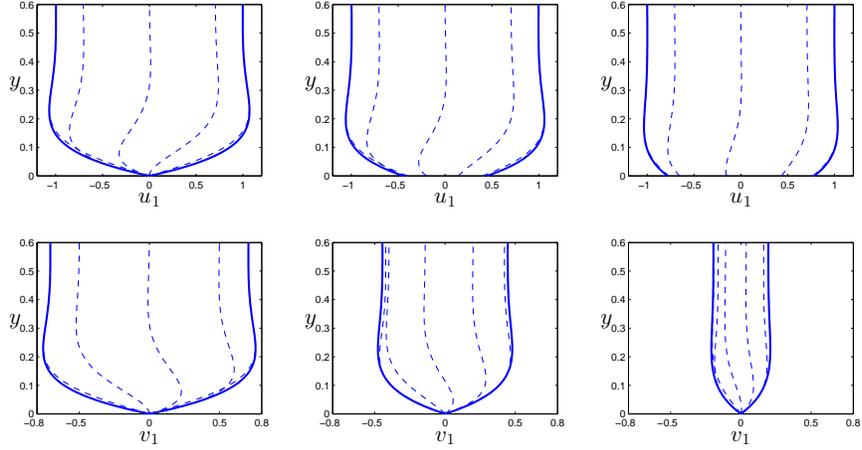}
\caption{ 
Wave field in the boundary layer.
The solid lines show the amplitude of the velocities $|u_1/U|$ (top row) and  $|v_1/U'|$ (bottom row) for $\beta=0$ (left), $0.5$ (middle) and $2$ (right). The time evolution is illustrated by the dashed lines showing $u_1/|U|$ and $v_1/|U'|$ at $x=0$ with assumption of a travelling wave $U\propto\e^{ \ii kx}$ and for the phases $\omega t =0,~ \pi/4,~ \pi/2,~ 3\pi/4,~ \pi$ (from right to left).}
\label{fig2}
\end{figure*}

It follows from (\ref{021}) and (\ref{014}) that $p_1$ is independent of $Y$, leading to
\begin{equation}
\partial_t u_1 =  \partial_t U_1 + \omega \partial^2_{YY} u_1/2.  \label{004}
\end{equation}
Solving (\ref{004}) with the boundary conditions
$u_1 \rightarrow U$ as $Y \rightarrow \infty $ and
$u_1 = \beta \partial u_1/\partial Y$ at $Y=0$,
we obtain
\begin{equation} 
u_1 = \Re \left(U  \e^{- \ii \omega t} \left(1-\frac{\e^{- (1- \ii) Y}}{1 + (1 -  \ii)\beta}\right)\right) \label{u1}
\end{equation}
to leading order in $k \delta$.
The equation of state implies that $p_1=c_0^2 \rho_1$ and, using (\ref{014}), that $\rho_1$ is independent  of $Y$: $\rho_1 =R_1$. Subtracting (\ref{005}) from  (\ref{021}), integrating and imposing $v_1$ bounded as $Y \to \infty$ then gives
\begin{equation} 
v_1 = \delta \Re \left(U'  \e^{- \ii \omega t}   \frac{(1 +  \ii)}{2(1 + (1 -  \ii)\beta)}\left(1-\e^{- (1 -  \ii) Y}\right) \right), \label{v1}
\end{equation}
also to leading order in $k \delta$. The two components $(u_1,v_1)$ of the wave velocity in the boundary layer for different values of $\beta$ are displayed in Figure \ref{fig2}. 
We only show the result of travelling wave, and the response to a standing wave is the same up to phase differences.
The figure indicates that the amplitude of the component $u_1$ of the wave velocity parallel to the wall is almost constant as $\beta$ varies  while the perpendicular component $v_1$ decreases as $\beta$ increases.

\section{Mean flow}

Using  the form (\ref{u1})--(\ref{v1}) for the wave field, we can calculate the Reynolds stress and solve the mean-flow equation which, in the boundary layer, takes the form
\begin{equation}
\omega \partial^2_{YY} \overline{u}_2 / 2 = \delta^{-1} \left(\partial_Y \overline{u_1 v_1} - \rbar{\partial_Y \overline{u_1 v_1}}_{\infty}\right) + \partial_x \overline{u_1^2}  - \rbar{\partial_x \overline{u_1^2}}_{\infty}, \label{016}
\end{equation}
where the subscripts $\infty$ indicate the limit $Y\to \infty$ and the overbars indicate averaging over a wave period. This expression is obtained by averaging (\ref{NSeq}), retaining only leading-order terms in $k \delta$, and subtracting from the inner equation its limit as $Y \to \infty$ to eliminate the $Y$-independent pressure term in exactly the same manner as employed for the wave equations. 

It is convenient to  consider the effect of $\partial_Y \overline{u_1v_1}$ and $\partial_x \overline{u_1 u_1}$ separately, taking advantage of the linearity of Eq.\ (\ref{016}) for $\overline{u}_2$. First we calculate the effect of $\partial_Y \overline{u_1v_1}$. A short computation leads to
\begin{widetext}
\begin{equation}
\begin{aligned}
\overline{u_1v_1} & = \frac{1}{8}\frac{ \delta  U {U'}^*}{(1+\beta)^2+\beta^2} \left( 1 -  \ii(1+2\beta) - (1 -  \ii)\e^{-(1 -  \ii)Y} \right. \\ &
 \left.- (1 -  \ii(1+2\beta))\e^{-(1+ \ii)Y} + (1- \ii) \e^{-2Y}  \right) +\cc,
\end{aligned}
\end{equation}
\end{widetext}
where $\cc$ denotes the complex conjugate of the preceding term.
Since $\partial_Y \overline{u_1 v_1} |_\infty = 0$, the $Y$-dependent terms immediately give the contribution to the shear $\omega \partial_Y \bar{u}_2/2$. 
Integrating these terms and using the averaged Navier boundary condition $\overline{u}_2 = \beta \partial_Y \overline{u}_2$ at $Y=0$ finally gives the first contribution to the mean velocity
\begin{widetext}
\begin{equation} \label{008}
\begin{aligned}
\overline{u}_2^{(\mathrm{i})} &= \frac{1}{4\omega }\frac{ U {U'}^*}{(1+\beta)^2+\beta^2} \left( \e^{-(1 -  \ii)Y}-1 + \frac{(1 -  \ii(1+2\beta))}{1+ \ii}(\e^{-(1+ \ii)Y}-1) \right. \\
& \left. - \frac{(1- \ii)}{2} (\e^{-2Y}-1)  
+ \beta (-1+ \ii(1+2\beta)) \right) + \cc . 
\end{aligned}
\end{equation}
\end{widetext}

\begin{figure*}
\centering
\includegraphics[width=14cm]{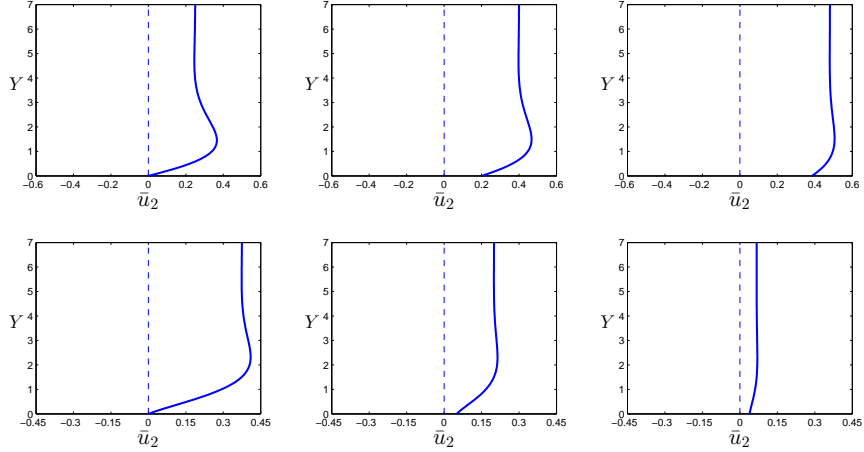}
\caption{Mean velocity profiles for the travelling wave  $U(x)=A \e^{ \ii kx}$ (top) and the standing 
wave $U(x)=A \cos(kx)$ (bottom) for $\beta=0$ (left), $0.1$ (middle) and $2$ (right). The profiles are normalized by $A^2/c_0$ (travelling wave) and $\sin(2kx)A^2/c_0$ (standing wave).}\label{mean}
\end{figure*}

Next we calculate the effect of $\partial_x \overline{u_1^2}$: starting with
\begin{widetext}
\begin{equation}
\omega \partial^2_{YY}  \overline{u}_2 /2= \partial_x \left(\overline{u_1^2} - \left. \overline{u_1^2} \right|_\infty\right) = - \frac{(|U|^2)'}{4} \left( \frac{2}{1 + \beta -  \ii\beta} \e^{- (1- \ii) Y} - \frac{\e^{-2Y}}{(1+\beta)^2+\beta^2} \right) +\cc, 
\end{equation}
\end{widetext}
integrating twice and applying the boundary conditions $\partial_Y \overline{u}_2 \to 0$ as $Y \to \infty$ and $\overline{u}_2 = \beta \partial_Y  \overline{u}_2$ at $Y=0$ yields
\begin{widetext}
\begin{equation}
\overline{u}_2^{(\mathrm{ii})} = \frac{1}{2 \omega}\frac{(|U|^2)'}{(1+\beta)^2+\beta^2} \left((\beta -  \ii (1+\beta))\e^{-(1- \ii)Y} + \frac{\e^{-2Y}}{4} - \frac{1}{4} - \frac{\beta}{2} \right) + \cc. \label{006}
\end{equation}
\end{widetext}
Combining (\ref{008}) and (\ref{006}) leads to the mean profile $\overline{u}_2=\overline{u}_2^{\mathrm{(i)}} + \overline{u}_2^{\mathrm{(ii)}}$. This is illustrated in Figure \ref{mean}
for a travelling wave with $U(x) = A \exp( \ii k x)$, and for a standing wave with $U(x)=A \cos(kx)$. 
As $\beta$ increases, the amplitude  of  $\overline{u}_2$ increases for the travelling wave and  decreases for the standing wave; we comment on the physical mechanism underlying this dependence in Sec.\  IV.

 Letting $Y \to \infty$ in $\overline{u}_2$, we obtain the total steady streaming velocity outside the boundary layer as
\begin{equation}
\overline{U}_2 = -\gamma^{\mathrm{s}}(|U|^2)'/\omega - \gamma^{\mathrm{t}}  \ii(U^* U'-U (U')^*)/\omega, \label{019}
\end{equation}
where
\[
\gamma^{\mathrm{s}} = \frac{3+4\beta}{8\left((1+\beta)^2+\beta^2\right)} \quad \textrm{and} \quad 
\gamma^{\mathrm{t}} = \frac{1+4\beta+4\beta^2}{8\left((1+\beta)^2+\beta^2\right)}.
\]
This expression provides an effective slip condition for the flow in the interior. It generalises to the Navier condition results obtained by \citet{Nybo1965} and \citet{Ligh1978, Ligh1978b} in the no-slip case $\beta=0$. Note that $(|U|^2)'= 2 \, \Re (U ^* U')$ and $ 
  \ii(U^* U'-U (U')^*) = - 2 \, \Im  (U ^* U')$ can be thought of as measuring the standing- and travelling-wave components of more general wave fields. 

We emphasise that (\ref{019}) gives the Eulerian mean flow:
results of this type can alternatively be  formulated in terms of the  Lagrangian mean slip velocity, as in Ref.\ \citep{Vann2011}. The difference between the two mean velocities is the Stokes drift, given outside the boundary layer by
\begin{equation}
\overline{U}^{\mathrm{Sto}}_2 = - \ii(U^* U'-U (U')^*)/(4\omega),
\end{equation}
leading to the Lagrangian mean slip velocity
\begin{eqnarray}
\overline{U}_2^\mathrm{L} = \overline{U}_2 + \overline{U}_2^{\mathrm{Sto}}  = -\gamma^{\mathrm{s}}(|U|^2)'/\omega - \left( \gamma^{\mathrm{t}} + 1/4 \right)  \ii(U^* U'-U (U')^*)/\omega \label{LagMe}
\end{eqnarray}
which may be more easily accessible in observations.

\begin{figure}
\centering
\includegraphics[width=8cm]{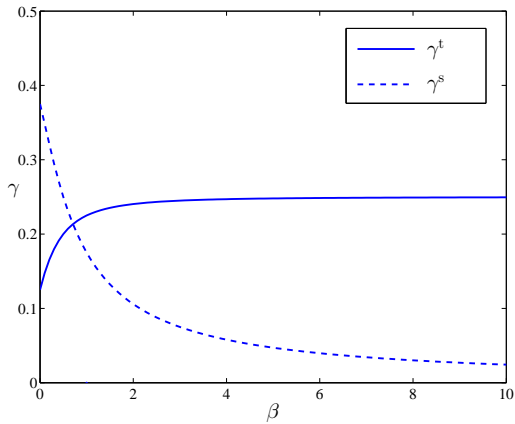}
\caption{Coefficients $\gamma^{\mathrm{t}}$ and $\gamma^{\mathrm{s}}$ in expression (\ref{019}) for the streaming velocity as a function of the slip parameter $\beta$. }
\label{fig1}
\end{figure}

From (\ref{019}) we can compute the steady streaming by travelling and standing waves, with $U=\hat U \exp( \ii k x)$ and $U(x)$ real, respectively, to find
\begin{equation}
\overline{U}_2^\mathrm{t} = 2\gamma^{\mathrm{t}} |\hat U|^2/ c_0 \quad \textrm{and} \quad
\overline{U}_2^\mathrm{s}= -2\gamma^{\mathrm{s}} U U'/\omega.
\end{equation}
These expressions, which provide an interpretation for the coefficients $\gamma^{\mathrm{t}}$ and $\gamma^{\mathrm{s}}$, reduce to well-known expressions  \citep[][Eq.\ (61)]{Nybo1965} and \citep[][Eq.\ (94)]{Ligh1978}, including  Rayleigh's result  for standing waves \citep{Rayl1896}, when $\beta=0$. The dependence of $\gamma^{\mathrm{t}}$ and $\gamma^{\mathrm{s}}$ on $\beta$ is illustrated in Figure \ref{fig1}. One (not necessarily intuitive) conclusion is that slip at the boundary increases the streaming velocity away from the boundary for travelling waves while it decreases the streaming velocity for standing waves. More specifically, in the limit of large slip $\beta \to \infty$, the streaming velocity for travelling waves is increased by a factor 2 for travelling waves but reduced to zero for standing waves. Correspondingly, the Lagrangian mean slip is increased by a factor 4/3 for travelling waves and tends to zero for standing waves. 

\section{Discussion}

This paper derives the general expression (\ref{019}) for the streaming velocity induced by  acoustic waves over a flat boundary with Navier boundary condition.
This expression can be used as an effective boundary condition for the mean flow in the interior when both interior and boundary streaming are important. Naturally, it reduces to well-known results in the no-slip case  $\beta=0$.

In the opposite limit $\beta\to \infty$, the two parameters $\gamma^{\mathrm{t}}$ and $\gamma^{\mathrm{s}}$ that appear in (\ref{019}) and are associated, respectively, with travelling and standing waves, behave very differently, with 
$\gamma^{\mathrm{t}} \to 1/4$ while $\gamma^{\mathrm{s}} \to 0$.
Physically,  the travelling wave  contribution $\gamma^{\mathrm{t}}$ stems from $\partial_Y\overline{u_1v_1}$, the divergence of the $y$-component of the flux of $x$-momentum, 
while the standing wave component contribution $\gamma^{\mathrm{s}}$ also stems  from $\partial_x\overline{u_1u_1}$, the divergence of the $x$-component of this flux. Only $u_1$  appears in $\partial_x\overline{u_1u_1}$; as $u_1$  becomes unaffected by the wall in the limit $\beta\to\infty$,  $\partial_x\overline{u_1u_1}$ clearly tends to $0$. The contribution of $\partial_Y\overline{u_1v_1}$ in the limit of $\beta\to\infty$, which involves the $y$-components of the wave fields, is more complex. Because the outer flow is fixed and the wave velocity is bounded, the wave shear, Reynolds stress and hence mean-flow forcing decrease as the slip length increases.
However, for a given mean-flow forcing, the mean velocity at the boundary and indeed across the boundary layer increases as the slip length increases (because the constraint imposed by the boundary condition weakens). The competition between the two effects leads to a balance as $\beta \to \infty$. Importantly, this shows the limit $\beta \to \infty$ to be singular, with the Navier condition yielding a different solution to a completely stress-free condition at the wall. The small but non-zero velocity perpendicular to the wall, $v_1 = O(\beta^{-1})$, imposed by mass conservation, leads to a non-zero mean momentum flux which in turns affects the mean flow in boundary layer at $O(1)$. This can be seen directly by combining the Navier boundary condition with the mean momentum equation to obtain
$
\overline{u}_2|_0 = L_\mathrm{s} \rbar{\partial_y \overline{u}_2}_0 = -\rbar{(L_\mathrm{s}/\nu)\overline{u_1v_1}}_\infty = -(L_\mathrm{s}/\nu) \overline{ U_1 V_1}$, neglecting the  contribution of $\partial_x \overline{u_1^2}$ which is $O(\beta^{-1})$. 
In the outer region, 
$U_1 = \Re \left(U \e^{- \ii \omega t} \right)$ and $V_1 = \Re \left( \ii  \delta U' \e^{- \ii \omega t}/(2\beta) \right)$
as $\beta \to \infty$ (see (\ref{u1})--(\ref{v1})); furthermore, $\overline{U}_2 \sim \overline{u}_2|_0$ (since $\overline{u}_2$ becomes independent of $Y$ as expected in the stress-free limit)  so that $ \overline{U}_2 \sim  -  \ii (U^* U'-U (U')^*)/(4\omega)$, consistent with (\ref{019}). It is only for standing waves, for which $U$ and $U'$ are in phase, that this vanishes.

We conclude with two remarks. First, different wave frequencies lead to very different mean velocity profiles because of the dependence of the boundary-layer thickness on the frequency. One can therefore propose that acoustic waves with a rich, variable wave spectrum may provide a method for controlling the mean-velocity profile near a solid boundary. Second, the dependence of the mean velocity on the slip length suggests that acoustic streaming could be used for the (notoriously difficult) estimation of the slip lengths of various fluid-solid combinations. An experiment estimating the Lagrangian slip velocity $\overline{U}_2^\mathrm{L}$ by measuring the mean speed of tracer particles would make it possible to infer $\beta$ from (\ref{LagMe}) and, since  $L_\mathrm{s} = \beta\sqrt{2\nu/\omega}$, the slip length. Carrying out such an experiment over a range of frequencies would ensure a good accuracy. The frequencies $\omega$ should be chosen with $\beta$  of order one so that it depends substantially on $\omega$. For instance, in water, if $L_\mathrm{s} \sim 100$ nm, $\beta$ varies from $1$ to $3$ as $\omega$ varies from about $0.1$ to $1$ GHz. One difficulty may be to ensure that the tracer particles provide an accurate estimate of the Lagrangian slip velocity: their motion may be affected by interior streaming  \cite[e.g.][]{Vann2011} if they are not  confined sufficiently close to the wall, and by radiation pressure \cite[e.g.][]{Doin1994R}.

\section{Acknowledgements}
The authors thank the two anonymous reviewers for their constructive comments.
Jin-Han Xie acknowledges financial support from the Centre for Numerical Algorithms and Intelligent Software (NAIS).


\end{document}